\documentclass[aps,prb,reprint,superscriptaddress,nofootinbib,floatfix]{revtex4-2}

\usepackage{amsmath,amssymb,bm}
\usepackage{graphicx}
\usepackage{xcolor}
\usepackage{booktabs}
\usepackage{tikz}
\usetikzlibrary{arrows.meta,positioning,shapes.geometric,calc,fit,backgrounds}
\usepackage[colorlinks=true,citecolor=blue,linkcolor=blue,urlcolor=blue]{hyperref}

\newcommand{\dd}{\mathrm{d}}
\newcommand{\ii}{\mathrm{i}}
\newcommand{\Tr}{\mathrm{Tr}}
\newcommand{\BZ}{\mathrm{BZ}}
\newcommand{\QGT}{Q}
\newcommand{\metric}{g}
\newcommand{\berry}{\Omega}

\newcommand{\Hqp}{H_{\mathrm{QP}}}
\newif\ifdraftnotes \draftnotesfalse

\graphicspath{{paper_figs/}{figs/}{./}}

\begin{document}

\title{Quantum weight and low-loss EELS signatures of Wannier quantum geometry in black phosphorus}

\author{Vinayak M. Kulkarni}
\email{vmkphysimath@gmail.com}
\affiliation{Independent Researcher, Vienna, Austria}
\altaffiliation{Former affiliation: Theoretical Sciences Unit, Jawaharlal Nehru Centre for Advanced Scientific Research, Bengaluru 560064, India}

\author{Sharona Horta}
\email{Sharona.Horta@ist.ac.at}
\affiliation{Institute of Science and Technology Austria, Am Campus 1, 3400 Klosterneuburg, Austria}

\date{\today}

\begin{abstract}
Quantum geometry is now experimentally accessible in crystalline solids, with black phosphorus providing a key platform through polarization-resolved angle-resolved photoemission spectroscopy. We develop a first-principles framework that connects the momentum-resolved quantum metric of black phosphorus to a complementary bulk observable: the direction-resolved quantum weight measurable through low-loss electron energy-loss spectroscopy (EELS). A 32-band DFT--Wannier Hamiltonian is used to compute both single-band and occupied-manifold geometric quantities from analytic momentum derivatives. We show that the raw single-band quantum metric of the top valence band is not globally meaningful in the conventional cell because folding degeneracies and intra-valence near degeneracies produce true isolated-band singularities; masked maps and occupied-manifold projectors are therefore essential. Because semilocal PBE produces near-gap semimetallic pockets and spurious subgap interpolation features, we introduce an experimentally motivated restricted quantum weight $K_{ii}(\omega_c)$, which obeys the corresponding restricted Souza--Wilkens--Martin sum rule and is the appropriate quantity for low-loss EELS once the zero-loss region is excluded. The restricted in-plane quantum weight is nearly isotropic, $K_{zz}/K_{xx}=0.972\pm0.005$ (armchair/zigzag), despite the strong band-mass anisotropy and armchair-only absorption onset of black phosphorus. Orbital-resolved Hubbard--Hartree corrections leave the absolute quantum weights rigid at the sub-percent level while producing a small but resolved armchair-directed drift of $K_{zz}/K_{xx}$, approximately $+0.46\%$ per eV of $U$. These results identify low-loss EELS spectral moments as a practical probe of integrated quantum geometry in an anisotropic layered material.
\end{abstract}

\maketitle

\section{Introduction}

The geometry of quantum states is now recognized as a physical property
of Bloch electrons rather than merely a mathematical descriptor. For a
Bloch eigenstate $|u_{n\mathbf{k}}\rangle$, the quantum geometric tensor
(QGT) encodes both the quantum distance between neighboring states and
the geometric phase accumulated under adiabatic motion in momentum
space~\cite{ProvostVallee1980,Berry1984,Resta2011}. Its real part is the
quantum metric and its imaginary part is the Berry curvature. While the
Berry curvature is central to topological transport, anomalous Hall
responses, and Chern-band physics~\cite{Haldane2004}, the quantum metric
has become increasingly important in flat-band superconductivity,
optical response, orbital susceptibility, and correlated topological
matter~\cite{PeottaTorma2015,Wang2021}.

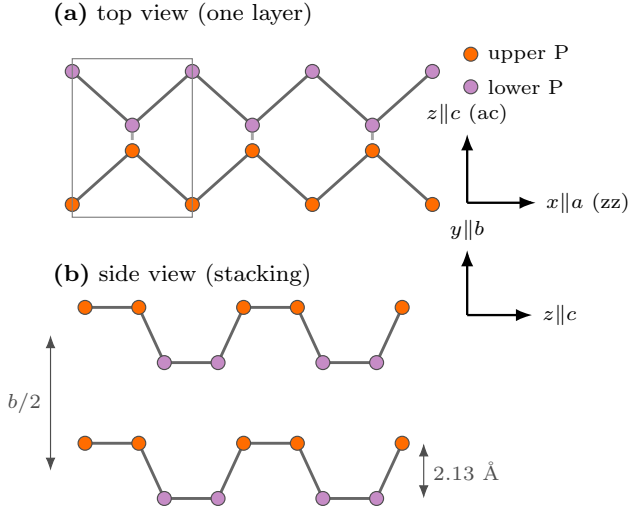
\begin{figure}[t]
\centering
\resizebox{0.98\columnwidth}{!}{%
\begin{tikzpicture}[scale=0.44,
  up/.style={circle, draw=black!70, fill=orange!85!red, inner sep=1.7pt},
  dn/.style={circle, draw=black!70, fill=violet!45, inner sep=1.7pt},
  b1/.style={line width=1.0pt, black!60},
  b2/.style={line width=1.0pt, black!35, dash pattern=on 2.4pt off 1.3pt},
  ax/.style={-{Latex[length=2.0mm]}, thick}]
\begin{scope}
\node[anchor=west] at (-0.8,5.6) {\footnotesize\textbf{(a)} top view (one layer)};
\foreach \i in {0,1,2}{
    \draw[b1] ({\i*3.3136},{0.353}) -- ({\i*3.3136+1.657},{1.836});
    \draw[b1] ({\i*3.3136+1.657},{1.836}) -- ({\i*3.3136+3.3136},{0.353});
    \draw[b1] ({\i*3.3136+1.657},{2.541}) -- ({\i*3.3136},{4.024});
    \draw[b1] ({\i*3.3136+1.657},{2.541}) -- ({\i*3.3136+3.3136},{4.024});
    \draw[b2] ({\i*3.3136+1.657},{1.836}) -- ({\i*3.3136+1.657},{2.541});
}
\foreach \i in {0,1,2,3}{
    \node[up] at ({\i*3.3136},{0.353}) {};
    \node[dn] at ({\i*3.3136},{4.024}) {};
}
\foreach \i in {0,1,2}{
    \node[up] at ({\i*3.3136+1.657},{1.836}) {};
    \node[dn] at ({\i*3.3136+1.657},{2.541}) {};
}
\draw[black!50, thin] (0,0) rectangle (3.3136,4.3763);
\draw[ax] (10.9,0.4) -- ++(1.9,0) node[right] {\scriptsize $x\|a$ (zz)};
\draw[ax] (10.9,0.4) -- ++(0,1.9) node[above] {\scriptsize $z\|c$ (ac)};
\node[up, label={right:\scriptsize upper P}] at (11.0,4.5) {};
\node[dn, label={right:\scriptsize lower P}] at (11.0,3.6) {};
\end{scope}
\begin{scope}[yshift=-8.6cm]
\node[anchor=west] at (-0.8,7.0) {\footnotesize\textbf{(b)} side view (stacking)};
\foreach \yc in {1.6, 5.35}{
  \foreach \j in {0,1}{
    \draw[b1] ({\j*4.3763+0.353},{\yc+0.762}) -- ({\j*4.3763+1.836},{\yc+0.762});
    \draw[b1] ({\j*4.3763+1.836},{\yc+0.762}) -- ({\j*4.3763+2.541},{\yc-0.762});
    \draw[b1] ({\j*4.3763+2.541},{\yc-0.762}) -- ({\j*4.3763+4.024},{\yc-0.762});
    \draw[b1] ({\j*4.3763+4.024},{\yc-0.762}) -- ({\j*4.3763+4.3763+0.353},{\yc+0.762});
  }
  \foreach \j in {0,1}{
    \node[up] at ({\j*4.3763+0.353},{\yc+0.762}) {};
    \node[up] at ({\j*4.3763+1.836},{\yc+0.762}) {};
    \node[dn] at ({\j*4.3763+2.541},{\yc-0.762}) {};
    \node[dn] at ({\j*4.3763+4.024},{\yc-0.762}) {};
  }
  \node[up] at ({2*4.3763+0.353},{\yc+0.762}) {};
}
\draw[{Latex[length=1.6mm]}-{Latex[length=1.6mm]}, black!70]
  (9.7,0.838) -- (9.7,2.362) node[midway, right] {\scriptsize $2.13$~\AA};
\draw[{Latex[length=1.6mm]}-{Latex[length=1.6mm]}, black!70]
  (-0.6,1.6) -- (-0.6,5.35) node[midway, left] {\scriptsize $b/2$};
\draw[ax] (10.9,5.9) -- ++(1.8,0) node[right] {\scriptsize $z\|c$};
\draw[ax] (10.9,5.9) -- ++(0,1.8) node[above] {\scriptsize $y\|b$};
\end{scope}
\end{tikzpicture}%
}
\caption{Crystal structure of bulk black phosphorus (Cmce, 8 atoms per
conventional cell), drawn to scale from the structural parameters used
in the calculations ($a=3.3136$, $b=10.473$, $c=4.3763$~\AA,
$u=0.10168$, $v=0.08056$)~\cite{BrownRundqvist1965}. (a)~Top view of
one puckered layer: zigzag ridges run along $x\|a$; upper (orange) and
lower (violet) pucker sublattices are connected by the cross-pucker
bond $d_2$ (dashed). (b)~Side view: AB-stacked layers separated by
$b/2$, layer~2 shifted by $a/2$ (C-centering); the pucker height is
$2ub = 2.13$~\AA.}
\label{fig:structure}
\end{figure}

A major recent development is the direct experimental access to the
quantum metric in crystalline solids. In black phosphorus,
polarization-dependent ARPES was used to reconstruct the momentum-space
pseudospin texture of the valence band and thereby obtain the full
quantum metric tensor~\cite{Kim2025Science,KimDryad2025}. Black
phosphorus is not an artificial photonic or cold-atom platform but a
real layered material with an anisotropic, puckered crystal structure
(Fig.~\ref{fig:structure}) and a narrow, highly tunable gap~\cite{Qiao2014,Low2014,Tran2014}. It
therefore provides a useful setting in which quantum-geometric band
theory, first-principles materials modeling, and momentum-resolved
spectroscopy can be compared.

Most current quantum-metric reconstructions, however, are based on
noninteracting band structures. In real materials the Bloch wave
functions are dressed by electronic interactions, phonons, disorder,
defects, strain, and substrate-induced hybridization. These effects can
modify the pseudospin texture even when the dispersion changes only
weakly; since the quantum metric depends on derivatives of the
normalized pseudospin field, small material-induced rotations of the
pseudospin can produce visible, anisotropic changes in the metric.

This motivates the central question of the present work:
\begin{quote}
\emph{How can one define, calculate, and experimentally test an
interaction-renormalized quantum geometry in a realistic anisotropic
narrow-gap material?}
\end{quote}

We answer it in three steps. First, a material-specific DFT--Wannier Hamiltonian for bulk black phosphorus supplies the band projectors and the velocity matrix elements entering the geometry. Second, a weak-coupling analysis identifies the leading static orbital channel generated by a short-range interaction; we implement this channel as an orbital-resolved Hubbard--Hartree self-energy on the full 32-band Wannier Hamiltonian. Third, we connect the resulting geometry to two complementary probes: polarization-resolved ARPES, which is momentum local and most naturally interpreted in an isolated two-band window~\cite{Kim2025Science}, and low-loss EELS, which accesses the direction-resolved occupied-manifold quantum weight through dielectric sum rules~\cite{SouzaWilkensMartin2000,OnishiFu2024qw}. This separation is essential in black phosphorus, where conventional-cell folding and intra-valence near degeneracies make raw single-band metrics singular while occupied-manifold quantities remain well defined.

The paper is organized as follows. Section~\ref{sec:workflow} presents
the theory--materials workflow. Section~\ref{sec:material_model}
defines the first-principles Wannier model and the effective two-band
description. Section~\ref{sec:qgt} formulates gauge-invariant quantum
geometry. Section~\ref{sec:interactions} introduces the interacting
Green's function, the weak-coupling shell analysis, and the self-consistent
Hartree implementation. Section~\ref{sec:arpes} connects to
polarization-resolved ARPES and Sec.~\ref{sec:eels} to quantum weight
and EELS. Section~\ref{sec:results} presents first-principles results
for bulk black phosphorus. Section~\ref{sec:discussion} discusses
limitations and extensions.

\begin{table}[t]
\centering
\caption{Axis conventions used throughout (conventional Cmce cell).
All computed tensors are reported in these Cartesian directions.}
\begin{tabular}{cccc}
\toprule
axis & lattice vector & length (\AA) & crystal direction \\
\midrule
$x$ & $a$ & 3.3136 & zigzag \\
$y$ & $b$ & 10.473 & stacking (vdW) \\
$z$ & $c$ & 4.3763 & armchair \\
\bottomrule
\end{tabular}
\label{tab:axes}
\end{table}

Throughout the paper, zigzag, stacking, and armchair denote three
inequivalent crystallographic directions of one bulk orthorhombic
black-phosphorus Hamiltonian. They should not be interpreted as three
separate lattices or three separate calculations. The band structure,
dielectric tensor, and quantum weights are all computed from the same
bulk 32-band Wannier model; $K_{xx}$, $K_{yy}$, and $K_{zz}$ are its
directional tensor components.

\section{Theory--materials workflow}
\label{sec:workflow}

\begin{figure}[t]
\centering
\resizebox{0.92\columnwidth}{!}{%
\begin{tikzpicture}[
  font=\scriptsize,
  >=Latex,
  box/.style={rectangle, rounded corners=2.3pt, draw=black!65, thick,
    align=center, minimum width=3.05cm, minimum height=0.78cm, inner sep=3pt},
  inputbox/.style={box, fill=blue!8},
  modelbox/.style={box, fill=orange!10},
  geobox/.style={box, fill=green!10},
  obsbox/.style={box, fill=purple!10},
  flow/.style={-{Latex[length=2.1mm]}, thick, draw=black!75},
  feedback/.style={-{Latex[length=1.8mm]}, thick, dashed, draw=black!55}
]

\node[inputbox] (mat)  at (0,0)     {\textbf{Material input}\\DFT structure\\lattice and axes};
\node[modelbox] (wan)  at (0,-1.12)  {\textbf{Wannier model}\\$H_0(\mathbf{k})$};
\node[modelbox] (self) at (0,-2.24)  {\textbf{Static self-energy}\\$\Sigma(0)$};
\node[geobox]   (geo)  at (0,-3.36)  {\textbf{Quantum geometry}\\$g_{ij}(\mathbf{k})$, $K_{ii}(\omega_c)$};
\node[obsbox]   (obs)  at (0,-4.48)  {\textbf{Observables}\\ARPES, low-loss EELS};
\node[modelbox, minimum width=2.35cm] (int) at (-3.10,-2.24)
  {\textbf{Interaction}\\short-range $U$};

\draw[flow] (mat) -- (wan);
\draw[flow] (wan) -- (self);
\draw[flow] (int) -- (self);
\draw[flow] (self) -- (geo);
\draw[flow] (geo) -- (obs);
\draw[feedback] (obs.east) -- ++(0.75,0) |- (mat.east);
\end{tikzpicture}%
}
\caption{Theory--materials workflow. DFT and structural information define a material-specific Wannier Hamiltonian. Short-range interactions feed a static orbital self-energy in the same Wannier basis. The resulting projectors define quantum-geometric observables that are compared with ARPES and low-loss EELS. The dashed feedback loop indicates how microscopy information about strain, defects, and interfaces can refine the starting Hamiltonian.}
\label{fig:workflow}
\end{figure}
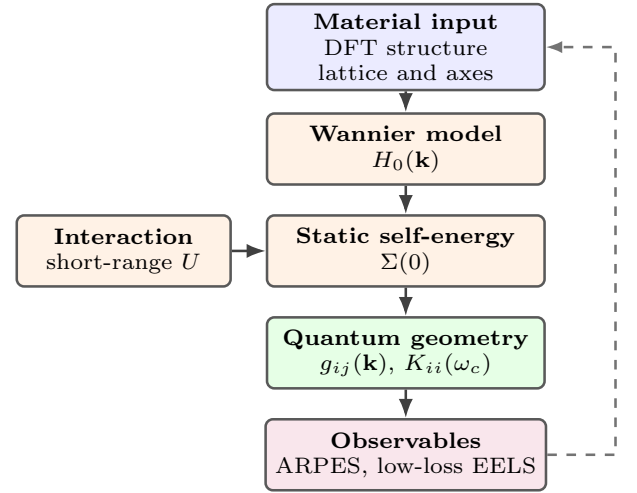

Figure~\ref{fig:workflow} summarizes the workflow. The material side
supplies a DFT-relaxed structure, orbital-resolved bands, and a Wannier
Hamiltonian for the low-energy subspace. The theory side converts this
Hamiltonian into gauge-invariant geometric observables, introduces
electronic interactions through a static orbital self-energy motivated by
the weak-coupling analysis of Sec.~\ref{sec:rg}, and predicts how the metric,
polarization-resolved ARPES maps, and EELS spectral moments change.

\section{Material-realistic model}
\label{sec:material_model}

\subsection{First-principles Wannier Hamiltonian}

The full low-energy model is a 32-band Wannier
Hamiltonian~\cite{MarzariVanderbilt2012,Wannier90_2020,Yates2007,Wang2006Wannier}
\begin{equation}
H_{mn}^{\mathrm{W}}(\mathbf{k}) = \sum_{\mathbf{R}}
t_{mn}(\mathbf{R})\, e^{\ii \mathbf{k}\cdot \mathbf{R}},
\label{eq:hw}
\end{equation}
constructed from $s$ and $p$ projections on the 8 phosphorus atoms of
the conventional cell (computational details in
Sec.~\ref{sec:compdetails}). The 20 valence bands and the conduction
edge are contained in the frozen disentanglement window and are
therefore reproduced exactly on the \textit{ab initio} grid.

\subsection{Effective two-band description}

Near the direct gap, the valence and conduction edge can be represented
by an effective two-band Hamiltonian
\begin{equation}
H_0(\mathbf{k}) = d_0(\mathbf{k})\tau_0 +
\mathbf{d}(\mathbf{k})\cdot \bm{\tau},
\label{eq:h0}
\end{equation}
with bands $\varepsilon_\pm = d_0 \pm |\mathbf{d}|$ and normalized
pseudospin $\hat{\mathbf{d}} = \mathbf{d}/|\mathbf{d}|$. For black
phosphorus such two-band descriptions can be fitted to tight-binding
models~\cite{Rudenko2014} or, as here, downfolded from
Eq.~\eqref{eq:hw}. Two caveats, quantified in Sec.~\ref{sec:results},
delimit its validity: (i) the top valence band approaches its
\emph{intra-valence} neighbor along a ring around $\Gamma$, and (ii)
bands of the conventional cell are exactly degenerate by folding on the
$k_x = \pm\pi/a$ zone face. Where either occurs, single-band and
two-band geometric quantities are ill-defined and the occupied-manifold
formulation of Sec.~\ref{sec:eels} must be used.

\section{Gauge-invariant quantum geometry}
\label{sec:qgt}

The QGT of a nondegenerate band $n$ is
\begin{equation}
\QGT_{ij}^{(n)}(\mathbf{k}) =
\langle \partial_i u_{n\mathbf{k}} |
\left(1-|u_{n\mathbf{k}}\rangle\langle u_{n\mathbf{k}}|\right)
| \partial_j u_{n\mathbf{k}}\rangle,
\label{eq:qgt_def}
\end{equation}
with $\metric_{ij} = \mathrm{Re}\,\QGT_{ij}$ and
$\berry_{ij} = -2\,\mathrm{Im}\,\QGT_{ij}$. Numerically we use the
gauge-invariant projector forms
\begin{align}
\metric_{ij}^{(n)} &= \tfrac{1}{2}\Tr\!\left[\partial_i P_n\,
\partial_j P_n\right],
\label{eq:metric_projector}\\
\berry_{ij}^{(n)} &= -\ii\Tr\!\left[P_n\,[\partial_i P_n,
\partial_j P_n]\right],
\label{eq:berry_projector}
\end{align}
with $P_n = |u_{n\mathbf{k}}\rangle\langle u_{n\mathbf{k}}|$, evaluated
from the analytic derivative $\partial_i H^{\mathrm W}(\mathbf k) =
\sum_{\mathbf R} \ii R_i\, t(\mathbf R)\, e^{\ii \mathbf k \cdot
\mathbf R}$ through the perturbation sum, avoiding finite-difference
stencils entirely~\cite{Yates2007,Wang2006Wannier}. For the two-band model,
Eqs.~\eqref{eq:metric_projector}--\eqref{eq:berry_projector} reduce to
\begin{align}
\metric_{ij} &= \tfrac{1}{4}\,\partial_i\hat{\mathbf{d}}\cdot
\partial_j\hat{\mathbf{d}},
\label{eq:metric_pseudospin}\\
\berry_{ij} &= -\tfrac{1}{2}\,\hat{\mathbf{d}}\cdot
\left(\partial_i\hat{\mathbf{d}}\times \partial_j\hat{\mathbf{d}}\right).
\label{eq:berry_pseudospin}
\end{align}

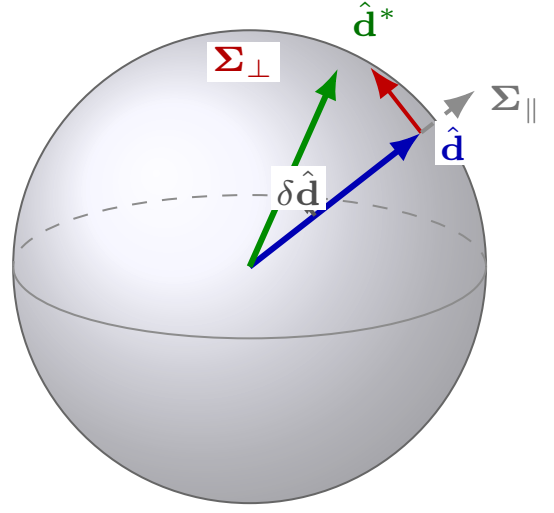
\begin{figure}[t]
\centering
\resizebox{0.82\columnwidth}{!}{%
\begin{tikzpicture}[>={Latex[length=2.2mm]}, every node/.style={font=\scriptsize}]
\shade[ball color=blue!8, opacity=0.55] (0,0) circle (1.65);
\draw[black!60] (0,0) circle (1.65);
\draw[black!45, dashed] (1.65,0) arc (0:180:1.65 and 0.50);
\draw[black!45] (-1.65,0) arc (180:360:1.65 and 0.50);

\draw[->, very thick, blue!70!black] (0,0) -- (38:1.52) coordinate (dtip);
\node[fill=white, inner sep=1pt, text=blue!70!black] at (1.42,0.86)
  {$\hat{\mathbf d}$};

\draw[->, very thick, green!55!black] (0,0) -- (66:1.52) coordinate (dstip);
\node[fill=white, inner sep=1pt, text=green!45!black] at (0.88,1.73)
  {$\hat{\mathbf d}^{*}$};

\draw[->, thick, red!75!black] (dtip) -- ++(128:0.58) coordinate (perpend);
\node[fill=white, inner sep=1pt, text=red!70!black, anchor=east] at (0.22,1.43)
  {$\bm{\Sigma}_{\perp}$};

\draw[->, thick, black!45, dashed] (dtip) -- ++(38:0.48) coordinate (parallel);
\node[fill=white, inner sep=1pt, text=black!55] at (1.86,1.13)
  {$\bm{\Sigma}_{\parallel}$};

\draw[->, thick, black!65] (38:0.58) arc (38:66:0.58);
\node[fill=white, inner sep=1pt, text=black!70] at (0.34,0.55)
  {$\delta\hat{\mathbf d}$};
\end{tikzpicture}%
}
\caption{Pseudospin renormalization on the Bloch sphere. Decomposing
the static self-energy as $\Sigma = \Sigma_0\tau_0 +
\bm{\Sigma}\cdot\bm{\tau}$, only the component
$\bm{\Sigma}_\perp$ transverse to the bare pseudospin rotates
$\hat{\mathbf d}$ and hence renormalizes the quantum geometry
[Eq.~\eqref{eq:deltad}]; the parallel component and the scalar part
shift the dispersion only.}
\label{fig:blochsphere}
\end{figure}

\section{Interaction-renormalized geometry}
\label{sec:interactions}

\subsection{Interacting Green's function and quasiparticle geometry}

Interactions enter through the matrix Green's function
\begin{equation}
G^{-1}(\mathbf{k},\omega)=\omega+\mu-H_0(\mathbf{k})-\Sigma(\mathbf{k},\omega).
\label{eq:green}
\end{equation}
For a system with well-defined quasiparticles we define
\begin{equation}
\Hqp(\mathbf{k}) = H_0(\mathbf{k}) + \mathrm{Re}\,\Sigma(\mathbf{k},0) - \mu,
\label{eq:hqp}
\end{equation}
decomposed in the two-band subspace as $\Hqp = d_0^*\tau_0 +
\mathbf{d}^*\cdot\bm{\tau}$, giving the renormalized pseudospin
$\hat{\mathbf{d}}^* = \mathbf{d}^*/|\mathbf{d}^*|$ and quasiparticle
metric $\metric^{\mathrm{QP}}_{ij} =
\tfrac14 \partial_i \hat{\mathbf d}^* \cdot \partial_j \hat{\mathbf
d}^*$. To linear order,
\begin{equation}
\delta\hat{\mathbf{d}}=
\frac{\bm{\Sigma}-\hat{\mathbf{d}}
(\hat{\mathbf{d}}\cdot\bm{\Sigma})}{|\mathbf{d}|}
+O(\bm{\Sigma}^2),
\label{eq:deltad}
\end{equation}
so a scalar self-energy leaves the geometry invariant while transverse
vector components rotate it (Fig.~\ref{fig:blochsphere}). Related
Green's-function-based constructions of interacting quantum metrics
have been analyzed for Hubbard models in Ref.~\cite{Sukhachov2025}; the
present work applies this principle to a material-specific Wannier
Hamiltonian.

\subsection{Weak-coupling limit: momentum-shell flow}
\label{sec:rg}

To motivate the form of the static self-energy, we consider a weak-coupling momentum-shell analysis of the
anisotropic band-edge model
\begin{equation}
\mathbf{d}(\mathbf{k}) =
\bigl(v_x k_x,\; v_z k_z,\; m - t_x k_x^2 - t_z k_z^2\bigr),
\label{eq:avec}
\end{equation}
with $v_x \neq v_z$ encoding the zigzag/armchair anisotropy
(Table~\ref{tab:axes}) and on-site Hubbard repulsion
$H_{\mathrm{int}} = U \sum_i n_{i\uparrow} n_{i\downarrow}$.
Integrating the momentum shell $\Lambda/b < |\mathbf{k}'| < \Lambda$ in
isotropized variables ($\bar v = \sqrt{v_x v_z}$) and rescaling gives
the tree-level dimensions at the band-touching fixed point,
\begin{equation}
[m] = 1, \qquad [v] = 0, \qquad [t] = -1, \qquad [U] = -1 ,
\end{equation}
i.e., the contact interaction is \emph{irrelevant} in $d=2$: with
$\tilde u \equiv U\Lambda/4\pi\bar v$, one has $d\tilde u/d\ell =
-\tilde u + \mathcal{O}(\tilde u^2)$. Interaction effects are therefore
finite, cumulative renormalizations of the relevant coupling $m$ ---
precisely the regime in which geometric observables acquire computable
corrections.

At one loop, for a pure contact interaction the exchange diagram
between opposite spins reduces to a density--density contraction and
the velocities are not renormalized,
$dv/d\ell = \mathcal{O}(\tilde u^2)$. The surviving diagram is the
Hartree bubble on the shell: each filled valence state carries
staggered density $\langle\sigma_z\rangle_{\mathbf k'} = -\hat
d_z(\mathbf k')$, so
\begin{equation}
\delta m = -\frac{U}{2}\int_{\text{shell}}
\frac{d^2k'}{(2\pi)^2}\,\hat d_z(\mathbf{k}'),
\label{eq:hartreeshell}
\end{equation}
yielding, with the tree-level rescaling,
\begin{align}
\frac{d\tilde m}{d\ell} &= \tilde m -
\tilde u\,\bigl\langle \hat d_z(\Lambda,\theta)\bigr\rangle_\theta
+ \mathcal{O}(\tilde u^2),
\label{eq:mflow}\\
\frac{d\tilde u}{d\ell} &= -\tilde u,
\qquad
\frac{d\tilde t}{d\ell} = -\tilde t .
\label{eq:uflow}
\end{align}
Because $\tilde u$ is irrelevant, the flow of $m$ terminates at a finite value. The shell analysis therefore identifies the leading static channel generated by a short-range interaction, but it should not be read as a complete treatment of all many-body effects. Our orbital-resolved Hartree calculation is the Wannier-level implementation of this leading weak-coupling channel. Dynamical correlations, vertex corrections, and any double-counting correction relative to the DFT starting point are beyond the present model. All interaction effects included below enter the geometry through the resulting static self-energy and the corresponding projectors.

\subsection{Orbital-resolved Hartree implementation on the Wannier model}
\label{sec:hartree_impl}

On the full 32-band Hamiltonian, symmetry forbids a
sublattice-staggered Hartree field (all P atoms belong to one Wyckoff
orbit), and the channel identified by Eq.~\eqref{eq:hartreeshell}
appears as an \emph{orbital-resolved} on-site term: in the paramagnetic
Hartree decoupling each Wannier orbital $o$ acquires
\begin{equation}
\Sigma_o = \frac{U}{2}\left(n_o - \bar n\right),
\label{eq:sigmaorb}
\end{equation}
with $n_o$ the self-consistent occupation of orbital $o$ (both spins)
and the orbital average $\bar n$ subtracted to pin the chemical
potential. Since the gap-edge states carry a specific orbital
polarization, the self-consistent solution of
Eq.~\eqref{eq:sigmaorb} renormalizes the crystal-field splittings, the
gap, and through them the quantum geometry. We emphasize the honest
labeling: this is a model coupling $U$ on the Wannier basis with no
double-counting correction against the underlying DFT. The shell analysis
only motivates the leading static channel; it does not make the calculation
a complete many-body treatment.

\subsection{Spectral geometry reconstructed from ARPES}

ARPES measures the matrix-element-weighted spectral function
$A(\mathbf{k},\omega) = -\pi^{-1}\mathrm{Im}\,G^R(\mathbf{k},\omega)$.
A spectroscopically reconstructed pseudospin over an energy window
$W$ is
\begin{equation}
\mathbf{s}_{W}(\mathbf{k})=
\frac{\int_{W}\dd\omega\,\Tr\left[\bm{\tau}
A(\mathbf{k},\omega)\right]}
{\int_{W}\dd\omega\,\Tr\left[A(\mathbf{k},\omega)\right]},
\label{eq:spectral_pseudospin}
\end{equation}
and the spectral metric is
$\metric^{\mathrm{spec}}_{ij} = \tfrac14 \partial_i
\hat{\mathbf{s}}_{W} \cdot \partial_j
\hat{\mathbf{s}}_{W}$. This is the experimentally relevant
metric associated with the pseudospin actually reconstructed from
spectral weight; it coincides with $\metric^{\mathrm{QP}}$ for sharp
quasiparticles and deviates when incoherent weight matters.

\section{Connection to polarization-resolved ARPES}
\label{sec:arpes}

Polarization-dependent intensities are schematically
$I_\lambda(\mathbf{k},\omega)\propto |M_\lambda(\mathbf{k})|^2
f(\omega) A(\mathbf{k},\omega)$; in favorable two-band situations,
combinations of linear and circular polarizations reconstruct the
pseudospin texture, as demonstrated for black
phosphorus~\cite{Kim2025Science,KimDryad2025}. The theoretically robust
comparisons are the gauge-invariant projector metric
[Eq.~\eqref{eq:metric_projector}] and the spectral metric
[Eq.~\eqref{eq:spectral_pseudospin}]; useful observables are the linear
and circular dichroism maps and window-integrated polarization ratios.
A conservative first treatment reuses the matrix-element model that
reproduces the noninteracting pseudospin map and inserts the
renormalized spectral function.

\section{Quantum weight and momentum-resolved EELS}
\label{sec:eels}

ARPES reconstructs $\hat{\mathbf d}(\mathbf k)$ locally but is
surface-sensitive and (quasi-)two-band. For bulk multiband systems a
complementary, fully gauge-invariant probe is the dielectric response.
Defining the rank-$N$ occupied projector
$P(\mathbf k) = \sum_{n\in\mathrm{occ}}
|u_{n\mathbf k}\rangle\langle u_{n\mathbf k}|$ and
\begin{equation}
G^{\mathrm{occ}}_{ij}(\mathbf k) = \tfrac{1}{2}\,\mathrm{Re}\,
\Tr\!\left[\partial_{i} P\, \partial_{j} P\right],
\label{eq:occmetric}
\end{equation}
the directional quantum weight
\begin{equation}
K_{ii} = \int_{\BZ} \frac{d^d k}{(2\pi)^d}\,
G^{\mathrm{occ}}_{ii}(\mathbf k)
\;\propto\; \int_0^{\infty} d\omega\,
\frac{\mathrm{Re}\,\sigma_{ii}(\omega)}{\omega}
\label{eq:swm}
\end{equation}
is fixed by the SWM sum rule~\cite{SouzaWilkensMartin2000} and is
experimentally accessible: momentum-resolved low-loss EELS measures the
loss function $\mathrm{Im}[-1/\varepsilon(\mathbf q,\omega)]$, from
which the absorptive dielectric function --- whose zeroth frequency
moment equals $4\pi^2 e^2 g_s K_{ii}$ in our conventions --- is
obtained by Kramers--Kronig analysis, following standard EELS methodology~\cite{Egerton2011} and established low-loss EELS band-structure applications~\cite{Salutari2024}. In 3D Coulomb systems the raw
loss moment and the quantum weight differ by screening (spectral weight
transferred to the plasmon)~\cite{OnishiFu2024qw}; the comparison must
therefore be made at the level of $\mathrm{Im}\,\varepsilon$, and the
difference between the two moments is itself a measure of screening.

Two features make $K_{ii}$ well suited here. First,
Eq.~\eqref{eq:occmetric} involves the occupied manifold as a whole:
intra-valence near-degeneracies cancel identically, so $K_{ii}$ is well
defined where single-band quantities are not. Second, the observable is
direction resolved: with $\mathbf q$ along the armchair or zigzag axis
(small-$q$ dipole limit), EELS measures the corresponding diagonal
component, and the ratio $K_{zz}/K_{xx}$ (armchair/zigzag,
Table~\ref{tab:axes}) is robust against normalization, thickness, and
zero-loss-subtraction uncertainties. The interaction flow of
Sec.~\ref{sec:rg} modifies $P(\mathbf k)$ near the gap edge and
produces a computable, $\Delta K_{ii}$.

\section{First-principles results for bulk black phosphorus}
\label{sec:results}

\subsection{Wannier model quality and the PBE semimetal}

The 32-band model reproduces the DFT valence manifold and conduction
edge exactly by construction (frozen window through the conduction
edge); the Wannier-grid gaps equal the DFT values to four decimals.
Wannier spreads are $1.45$~\AA$^2$ ($s$-like) and
$2.18$--$2.49$~\AA$^2$ ($p$-like), and the largest hoppings at the
outermost real-space shell are $1.2\times10^{-5}$ of the on-site scale
on the production $10\times8\times10$ anchor grid.

Two findings about the PBE reference itself emerged from this construction and are of general methodological interest. First, dense Brillouin-zone sampling exposes near-gap pockets that are missed on sparse Wannier grids; consequently, apparent PBE gap values depend strongly on the sampling used to diagnose them. Second, these near-gap pockets dominate unrestricted quantum-weight integrals because the occupied--unoccupied denominator enters quadratically. We therefore use a scissor-corrected reference as a low-energy regularization of the conduction manifold and define the experimentally relevant restricted weight of Sec.~\ref{sec:qw_results}. The scissor does not by itself rotate the wave functions and hence does not create a geometric renormalization; it changes energy denominators and separates the physically relevant interband spectral weight from the numerically fragile subgap region. All quantitative geometric tables below use restricted integrals that exclude the energy window where the semilocal gap error, zero-loss subtraction, and finite-range Wannier interpolation artifacts dominate.

\subsection{Single-band geometry and its domain of validity}

\begin{figure}[t]
\centering
\includegraphics[width=\columnwidth]{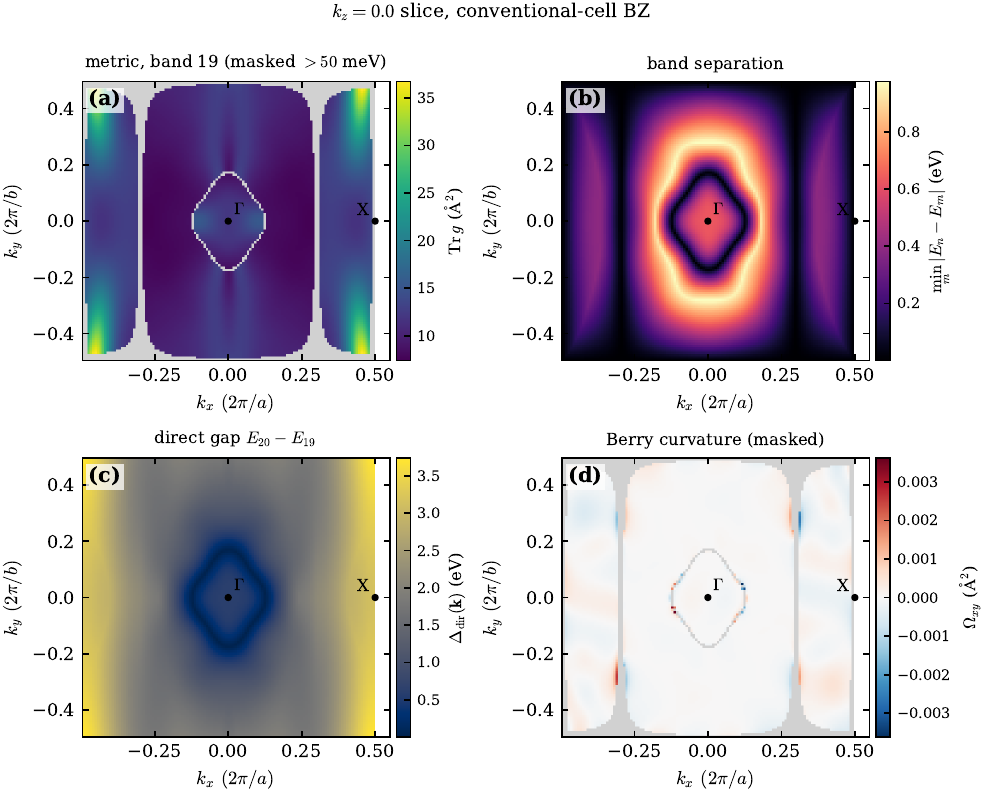}
\caption{Single-band quantum geometry of the top valence band on the
$k_z=0$ plane of the conventional-cell BZ. (a)~Quantum-metric trace,
masked (gray) where the band separation is below $50$~meV; (b)~minimum
band separation, showing the folding planes at $k_x=\pm\pi/a$ and the
intra-valence near-degeneracy ring around $\Gamma$; (c)~direct gap
$E_{20}(\mathbf{k})-E_{19}(\mathbf{k})$; (d)~Berry curvature
$\Omega_{xy}$ (masked), vanishing in the interior as required by
inversion and time-reversal symmetry. The $50$-meV mask removes both
the physical degeneracies and the interpolation-artifact pockets of
Sec.~\ref{sec:qw_results}.}
\label{fig:qgtmaps}
\end{figure}

Figure~\ref{fig:qgtmaps} shows the masked metric of the top valence
band on the $k_z = 0$ plane together with the minimum band-separation
map. Three structures organize the plane: the exact folding
degeneracies of the conventional cell on $k_x = \pm\pi/a$, an
intra-valence near-degeneracy ring around $\Gamma$ (separations below
$50$~meV), and the regular interior, where the masked metric reaches
$\mathrm{Tr}\,g \approx 34$~\AA$^2$. On the degenerate sets the
single-band metric genuinely diverges; all single-band maps are
therefore masked at $50$~meV separation, and integrated statements are
made with the occupied-manifold quantities below. This quantifies the
warning of Sec.~\ref{sec:material_model}: the effective two-band
description of the ARPES experiment is valid in the interior region
and fails on the ring, where a rank-2 valence-manifold treatment is
required.

\subsection{Restricted directional quantum weight}
\label{sec:qw_results}

The unrestricted quantum weight of the PBE-level Hamiltonian is not a
well-posed Wannier-representable quantity: PBE places bulk black
phosphorus close to an indirect semimetallic regime with direct gaps
collapsing to a few meV in small Brillouin-zone pockets. The resulting
near-singular conduction projector is difficult to represent faithfully
with a finite real-space Wannier truncation, and the interpolated gap
can dip spuriously between the \textit{ab initio} anchor points. We
therefore compute the \emph{energy-restricted} quantum weight
\begin{equation}
K_{ii}(\omega_c) = \int_{\BZ}\frac{d^3k}{(2\pi)^3}
\sum_{\substack{n\in\mathrm{occ},\,m\notin\mathrm{occ}\\
E_{m\mathbf k}-E_{n\mathbf k}>\omega_c}}
\frac{\bigl|\langle n|\partial_{k_i}H|m\rangle\bigr|^2}
{(E_{m\mathbf k}-E_{n\mathbf k})^{2}},
\label{eq:Kres}
\end{equation}
which obeys the corresponding restricted sum rule
$\int_{\omega_c}^{\infty}\mathrm{Im}\,\varepsilon_{ii}\,d\omega =
4\pi^2 e^2 g_s K_{ii}(\omega_c)$ exactly, pair by pair, for the same
Hamiltonian and cutoff. This is also the relevant low-loss EELS object:
spectral weight below $\omega_c\sim0.15$~eV is dominated by the
zero-loss region and by the numerical subgap sector rather than by the
robust interband response. The restriction is therefore not an ad hoc
fit parameter but a common lower integration limit applied to the
projector expression and to the absorptive dielectric spectrum. In
practice $K_{ii}(\omega_c)$ forms a double plateau, varying by only
$\sim\!1\%$ over $\omega_c=0.15$--$0.25$~eV and converging to
$\leq0.3\%$ across Brillouin-zone grids from $9^3$ to $21^3$ with
adaptive refinement (Table~\ref{tab:conv_wc}). The restricted weight,
not the unrestricted near-gap quantity, is therefore the experimental
observable emphasized below.

The resulting weights (Table~\ref{tab:qw}) show an unexpected
near-isotropy of the in-plane integrated response: although black phosphorus is the
canonical anisotropic layered semiconductor --- with strongly
direction-dependent masses and the armchair-only absorption onset
reproduced in Fig.~\ref{fig:eels} --- the frequency-integrated
geometric weight is nearly identical along armchair and zigzag,
$K_{zz}/K_{xx} = 0.972 \pm 0.005$. The dichroism of the onset is
compensated by zigzag-polarized weight at higher energies, such that
the integrated in-plane geometry is balanced to within $3\%$. This
near-equality of two integrated loss moments is directly testable by
$q$-resolved EELS with $\mathbf q$ along the two in-plane axes,
independent of absolute normalization. The stacking-direction weight
$K_{yy}$ is the largest component ($K_{xx}/K_{yy} = 0.255$), subject to
the position-operator caveat of Sec.~\ref{sec:discussion}.

\begin{table}[t]
\centering
\caption{Restricted directional quantum weight of bulk black
phosphorus (\AA$^{-1}$; scissor-corrected Hamiltonian, $21^3$ grid with
adaptive refinement, $\omega_c = 0.20$~eV). Parenthetical uncertainties combine the
grid spread ($9^3$--$21^3$) and the cutoff spread
($\omega_c = 0.15$--$0.25$~eV); they are strongly \emph{correlated}
between rows (identical grids and cutoffs), so differences between
rows are determined to $\sim\!0.001$ --- the interaction drift of the
ratio is resolved despite the row uncertainties.
$^\dagger$provisional pending the Wannier position-operator
correction.}
\resizebox{\columnwidth}{!}{%
\begin{tabular}{lccc|cc}
\toprule
 & $K_{xx}$ & $K_{yy}$ & $K_{zz}$ & $K_{zz}/K_{xx}$ & $K_{xx}/K_{yy}$ \\
 & (zigzag) & (stacking) & (armchair) & & \\
\midrule
$U=0$ & 0.1094(3) & 0.4278(11)$^\dagger$ & 0.1063(6) & 0.972(5) & 0.256(2) \\
$U=1$~eV & 0.1092(3) & 0.4283(11)$^\dagger$ & 0.1067(6) & 0.976(5) & 0.255(2) \\
$U=2$~eV & 0.1091(3) & 0.4287(11)$^\dagger$ & 0.1070(6) & 0.981(5) & 0.254(2) \\
\bottomrule
\end{tabular}%
}
\label{tab:qw}
\end{table}

\subsection{Dielectric response and geometric sum rule}

\begin{figure*}[t]
\centering
\includegraphics[width=\textwidth]{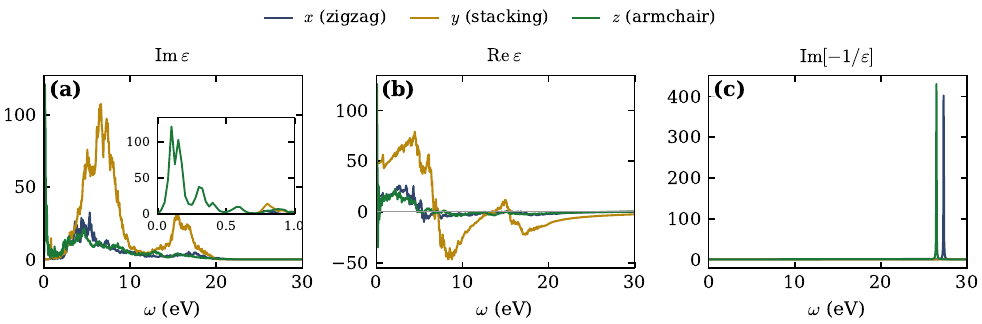}
\caption{Directional dielectric response of bulk black phosphorus from
the scissor-corrected Wannier model (independent-particle RPA,
small-$q$ dipole limit, $\eta=20$~meV). In this limit $\mathbf q$ along
$x$, $y$, or $z$ selects the corresponding longitudinal component of the
dielectric tensor. (a)~$\mathrm{Im}\,\varepsilon(\omega)$ with an inset on
the absorption onset: the near-gap transition is concentrated in the
armchair channel, reproducing the hallmark linear dichroism of black
phosphorus from first-principles matrix elements. (b)~$\mathrm{Re}\,\varepsilon(\omega)$.
(c)~Loss proxy $\mathrm{Im}[-1/\varepsilon]$; absolute plasmon peak
positions and heights are qualitative only because local-field effects,
excitons, finite-$q$ corrections, finite thickness, and plural scattering
are not included.}
\label{fig:eels}
\end{figure*}

Figure~\ref{fig:eels} shows the independent-particle dielectric function and the corresponding low-$q$ loss proxy for the three crystal directions. The calculated $\mathrm{Im}\,\varepsilon(\omega)$ reproduces the characteristic linear dichroism of black phosphorus: the near-onset spectral weight is concentrated in the armchair channel and is strongly suppressed for zigzag polarization. The relevant quantitative check is the restricted SWM relation between $K_{ii}(\omega_c)$ and the frequency integral of $\mathrm{Im}\,\varepsilon_{ii}$ above the same cutoff. In this restricted window the geometry computed directly from projectors and the absorptive dielectric response agree at the percent level. By contrast, the absolute plasmon peak positions and heights in $\mathrm{Im}[-1/\varepsilon]$ are qualitative in the present approximation because local-field effects, excitons, finite-$q$ corrections, finite sample thickness, and the complete high-energy conduction manifold are not included.

\subsection{Directional EELS anisotropy contrast}
\label{sec:eels_contrast}

The intrinsic quantity behind a momentum-resolved EELS experiment is the
longitudinal loss function
\begin{equation}
L_{\hat{\mathbf q}}(\omega)=
\mathrm{Im}
\left[-\frac{1}{\varepsilon_L(\hat{\mathbf q},\omega)}\right],
\qquad
\varepsilon_L(\hat{\mathbf q},\omega)=
\hat q_i\,\varepsilon_{ij}(\omega)\,\hat q_j .
\label{eq:directional_loss}
\end{equation}
In the small-$q$ dipole limit used here, choosing $\mathbf q\parallel x$
selects the zigzag response and choosing $\mathbf q\parallel z$ selects
the armchair response. Absolute experimental intensities additionally
depend on beam energy, convergence and collection angles, sample
thickness, plural scattering, and zero-loss subtraction~\cite{Egerton2011}.
For this reason the most robust theoretical signatures are normalized
ratios rather than absolute peak heights. We use the energy-resolved
contrast
\begin{equation}
A_{zx}(\omega)=
\frac{L_z(\omega)-L_x(\omega)}
     {L_z(\omega)+L_x(\omega)}
\label{eq:eels_contrast}
\end{equation}
and the integrated spectral-moment ratio
\begin{equation}
R_{zx}(\Omega)=
\frac{\int_{\omega_c}^{\Omega} d\omega\,\mathrm{Im}\,\varepsilon_{zz}(\omega)}
{\int_{\omega_c}^{\Omega} d\omega\,\mathrm{Im}\,\varepsilon_{xx}(\omega)} .
\label{eq:moment_ratio}
\end{equation}
When the upper window captures the relevant interband spectral weight,
$R_{zx}$ approaches the restricted quantum-weight ratio
$K_{zz}(\omega_c)/K_{xx}(\omega_c)$. Thus the predicted EELS signature
is twofold: a strongly armchair-polarized near-onset loss response,
followed by an integrated moment that becomes nearly isotropic in plane.

\subsection{Interaction renormalization: spectral softening with
geometric rigidity}
\label{sec:int_results}

Solving the orbital-resolved Hartree problem [Eq.~\eqref{eq:sigmaorb}]
self-consistently on the scissored Hamiltonian for $U = 1$ and $2$~eV
yields a symmetry-preserving solution (all eight P atoms equivalent;
occupation sum conserved to $10^{-4}$) with the negative-feedback sign
derived in Sec.~\ref{sec:rg}: the over-occupied $s$ orbitals are pushed
up and the $p$ manifold down, with $\max_o|\Sigma_o| = 0.24$ ($0.47$)~eV
at $U=1$ ($2$)~eV, and the orbital polarization is screened.

\begin{figure}[t]
\centering
\includegraphics[width=\columnwidth]{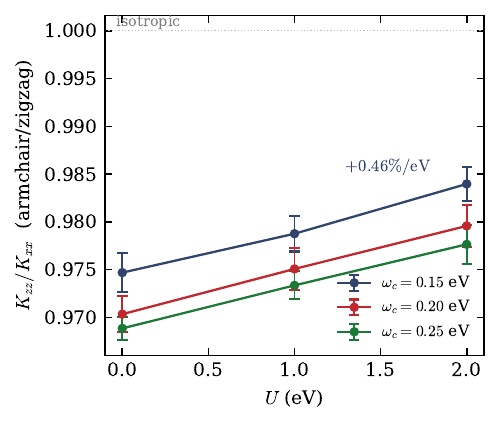}
\caption{Interaction flow of the in-plane quantum-weight anisotropy.
$K_{zz}/K_{xx}$ (armchair/zigzag) versus Hubbard $U$ for three
sum-rule cutoffs $\omega_c$; points are averages over Brillouin-zone
grids $9^3$--$21^3$ (adaptive refinement), error bars the half-spread
across grids. The slope $+0.46\%$~eV$^{-1}$ is identical across all
grids and cutoffs; the dotted line marks in-plane isotropy.}
\label{fig:ratioflow}
\end{figure}

\begin{figure}[t]
\centering
\includegraphics[width=\columnwidth]{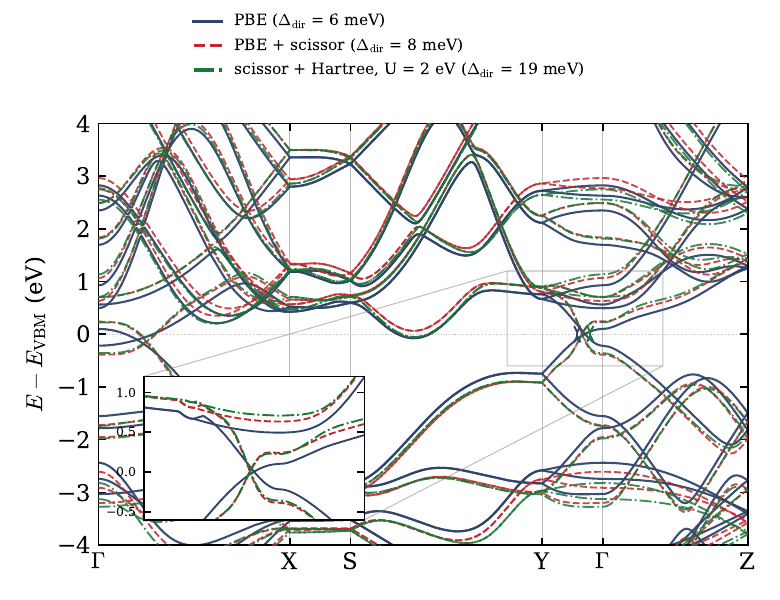}
\caption{Band structure of the same bulk 32-band black-phosphorus
Wannier Hamiltonian along the conventional-cell path
$\Gamma$--X--S--Y--$\Gamma$--Z, which samples the zigzag, stacking, and
armchair reciprocal directions. The three curves correspond to the three
Hamiltonians of this work --- bare PBE, quasiparticle (scissor), and
interaction-renormalized (Hubbard--Hartree at representative $U$) ---
each aligned to its own valence-band maximum, with the direct gap of each
Hamiltonian given in the legend, gap arrows marking its location on the
path, and a zoom inset on the gap region. The spurious conduction dip on the
$\Gamma$--Y segment of the PBE curve is the Wannier-interpolation
artifact analyzed in Sec.~\ref{sec:qw_results}. Valence bands of the PBE and scissored Hamiltonians
coincide exactly; folded bands of the conventional cell stick along the
X--S--Y boundary segments.}
\label{fig:bands}
\end{figure}

The response of the spectrum and of the geometry then separate
sharply. The \emph{indirect} gap softens strongly and nearly linearly,
by $-46$ and $-118$~meV at $U = 1$ and $2$~eV --- a linear extrapolation of the Hartree trend would close the
indirect gap near $U \sim 3.4$~eV, a scale to be read only as a model
diagnostic of the mean-field treatment, not as a quantitative
transition prediction --- while the
\emph{direct} gap, which controls the optical onset, is almost inert
($-0.6$ and $-3.1$~meV): at the direct-gap momentum the band-edge
states share nearly identical orbital composition and shift rigidly,
whereas the band extrema that define the indirect gap carry different
$s/p$ character and are pulled apart. The three band structures are compared in Fig.~\ref{fig:bands}.
The quantum weight inherits this
rigidity: $|\Delta K_{ii}/K_{ii}| \leq 0.7\%$ at $U = 2$~eV
(Table~\ref{tab:qw}), consistent with the weak-coupling irrelevance of the contact channel.
Within the rigidity, however, a
\emph{resolved} interaction fingerprint survives: the in-plane
anisotropy ratio rotates monotonically toward armchair,
\begin{equation}
\frac{d}{dU}\!\left(\frac{K_{zz}}{K_{xx}}\right)
\;\approx\; +0.46\%~\mathrm{eV}^{-1},
\label{eq:ratioflow}
\end{equation}
with identical slope (to two digits) across all Brillouin-zone grids
and all sum-rule cutoffs --- $24$ independent discretizations without a
single non-monotonic instance --- decomposing into
$\Delta K_{zz} = +0.65\%$ and $\Delta K_{xx} = -0.2\%$ at $U=2$~eV (Fig.~\ref{fig:ratioflow}).

The corresponding momentum-resolved EELS consequences are: an interaction-inert absorption
onset, a near-isotropic in-plane integrated loss moment, and a
sub-percent, armchair-directed drift of their ratio with interaction
strength.

\section{Experimental predictions}

The calculation leads to three experimentally testable low-loss EELS signatures in black phosphorus. First, the near-onset loss contrast $A_{zx}(\omega)$ should be strongly armchair polarized when $\mathbf q$ is rotated from zigzag to armchair. Second, the integrated in-plane spectral moment should approach a nearly isotropic value, $R_{zx}(\Omega)\to K_{zz}/K_{xx}\simeq0.97$, even though the onset is strongly armchair polarized. Third, weak orbital-selective interactions or material perturbations that mainly shift the indirect gap should leave the absolute restricted quantum weights nearly rigid, with the most sensitive fingerprint being a small monotonic drift of $K_{zz}/K_{xx}$ toward the armchair direction. Because these ratios compare two measurements performed on the same specimen, they are less sensitive to absolute intensity normalization, thickness, and zero-loss subtraction than either component separately.

\section{Discussion}
\label{sec:discussion}

The framework deliberately separates the noninteracting band metric,
the quasiparticle metric, the ARPES-reconstructed spectral metric, and
the occupied-manifold quantum weight, because no single definition of
an interacting quantum geometry is simultaneously gauge invariant,
directly measurable, and valid in all
regimes~\cite{Sukhachov2025,OnishiFu2024qw}. The material results of
Sec.~\ref{sec:results} add two practical lessons: quantum-geometric
observables are far more sensitive to Wannierization quality than band
structures, and cell/axis conventions must be fixed explicitly
(Table~\ref{tab:axes}) since black-phosphorus literature mixes several.

Limitations and controlled extensions of the present calculation are: the position-operator
correction to the Wannier velocity~\cite{Blount1962,Yates2007,Wang2006Wannier} (removing the residual
symmetry-forbidden $K_{yz}$ and firming up the stacking-direction
response); local-field and excitonic corrections to the out-of-plane
dielectric response; $k$-grid convergence of the quantum weights; and,
for the ARPES connection, photoemission matrix elements beyond the
model used to reproduce the noninteracting pseudospin map. On the
experimental side, the natural measurement is monochromated low-loss
STEM-EELS on black phosphorus with $\mathbf q$ along armchair and
zigzag, Kramers--Kronig analysis to $\mathrm{Im}\,\varepsilon$, and the
window-integrated moment ratio compared to $K_{zz}/K_{xx}$; the local
character of STEM additionally enables strain- and defect-resolved
quantum-weight mapping --- leveraging demonstrated atomic-plane-resolved
EELS~\cite{Negi2017} and interface-resolved STEM
characterization~\cite{Xu2025Science} --- an observable inaccessible
to far-field optics.

\section{Conclusion}

We developed a DFT--Wannier framework for connecting quantum geometry in black phosphorus to bulk low-loss EELS observables. The central technical point is that isolated-band quantum metrics and occupied-manifold quantum weights play different roles: the former are appropriate for ARPES-like momentum-local reconstructions only where the band is isolated, while the latter remain well defined across intra-valence degeneracies and control integrated dielectric spectral weight. Applying this framework to a 32-band Wannier model, we find a restricted in-plane quantum weight that is nearly isotropic, $K_{zz}/K_{xx}=0.972\pm0.005$, despite the strong armchair/zigzag anisotropy of the band dispersion and absorption onset. Orbital-resolved Hubbard--Hartree corrections produce sizable spectral shifts in the indirect gap but leave the restricted quantum weights rigid at the sub-percent level, with only a small monotonic drift of the in-plane ratio. These predictions can be tested by monochromated, momentum-resolved low-loss EELS with $\mathbf q$ along the armchair and zigzag directions, and the same workflow can be extended to strained, defective, and interface-engineered puckered materials and related layered systems.

\section*{Author contributions}

V.M.K. conceived and led the project, developed the theoretical framework,
formulated the quantum-weight and EELS connection, carried out the
projector-based quantum-geometry analysis, implemented the scissor and
orbital-resolved Hubbard--Hartree calculations, generated the numerical
tables and figures, and wrote the manuscript.

S.H. contributed the first-principles and materials side of the work,
including the construction and validation of the black-phosphorus crystal
structure and lattice conventions, DFT/Wannier materials input, and the
materials and EELS interpretation. Both authors discussed the results,
interpreted the physical implications, revised the manuscript, and
approved the final version.

\section*{Computational resources}

The numerical workflow reported here was carried out on personal
computing resources, primarily an Apple MacBook with Apple M2 processor.
This includes the Wannier-Hamiltonian analysis, quantum-geometric tensor
calculations, restricted quantum-weight integrations, scissor corrections,
Hubbard--Hartree iterations, dielectric post-processing, and figure
generation. No institutional high-performance-computing resources were
used for the calculations reported here.
\begin{acknowledgments}
V.M.K. acknowledges doctoral training at the Theoretical Sciences Unit,
Jawaharlal Nehru Centre for Advanced Scientific Research, Bengaluru.
S.H. acknowledges the research environment of the Ib{\'a}{\~n}ez group
at ISTA. The authors
acknowledge the developers of \textsc{Quantum ESPRESSO} and
\textsc{Wannier90}, and the open-source Python scientific-computing
ecosystem used for post-processing and visualization.
\end{acknowledgments}

\appendix

\section{Computational details}
\label{sec:compdetails}

DFT: \textsc{Quantum ESPRESSO}~\cite{QE2009}, PBE, ultrasoft
pseudopotential, $E_{\mathrm{cut}} = 60/480$~Ry, conventional Cmce cell
($a=3.3136$, $b=10.473$, $c=4.3763$~\AA, 8 P atoms at $8f$ with
$u=0.10168$, $v=0.08056$~\cite{BrownRundqvist1965}); SCF grid
$12\times8\times10$, NSCF/Wannier grid $10\times8\times10$ with 40
bands (a sparser $8\times4\times8$ grid was found to mask the PBE
band overlap and to under-determine the real-space Hamiltonian near the
gap; see Sec.~\ref{sec:qw_results}).
Wannierization: \textsc{Wannier90}~\cite{Wannier90_2020}, 32 functions
($s+p$ per atom), outer window $[-30,50]$~eV, frozen window through the
conduction edge ($[-30, E_{\mathrm{VBM}}+1.0]$~eV); spreads
$1.45$--$2.43$~\AA$^2$. Geometry: analytic $\partial_k H$ perturbation sums (no finite differences), degeneracy-masked single-band maps, and restricted occupied-manifold weights. Dielectric response: independent-particle RPA, $q\to0$, Lorentzian $\eta = 20$~meV; the restricted sum rule is checked above the same cutoff used in Eq.~\eqref{eq:Kres}. Scissor: rigid conduction-subspace shift in the Wannier representation, used as a low-energy regularization while preserving the relevant projectors. Quantum weights: restricted form Eq.~\eqref{eq:Kres}
with $\omega_c = 0.15$--$0.25$~eV, grids $9^3$--$21^3$ with adaptive
refinement (cells with occupied--unoccupied separation below $1.2$~eV
re-integrated on $5^3$ subgrids). Interaction:
self-consistent orbital-resolved Hartree, Eq.~\eqref{eq:sigmaorb},
linear mixing, occupation tolerance $10^{-7}$.

\section{Convergence of the restricted quantum weight}
\label{app:conv}

Table~\ref{tab:conv_wc} documents the double plateau of the restricted
armchair quantum weight $K_{zz}(\omega_c)$ of the scissor-corrected
Hamiltonian in both Brillouin-zone grid density and sum-rule cutoff;
$K_{xx}$ and $K_{yy}$ are stable at the $10^{-3}$ level over the same
ranges (not shown).

\begin{table}[h]
\centering
\caption{$K_{zz}$ (\AA$^{-1}$) of the scissor-corrected Hamiltonian
versus Brillouin-zone grid and sum-rule cutoff $\omega_c$ (adaptive
refinement throughout).}
\begin{tabular}{lccc}
\toprule
grid & $\omega_c=0.15$~eV & $0.20$~eV & $0.25$~eV \\
\midrule
$9^3$  & 0.1068 & 0.1060 & 0.1058 \\
$11^3$ & 0.1067 & 0.1062 & 0.1057 \\
$15^3$ & 0.1069 & 0.1063 & 0.1057 \\
$21^3$ & 0.1070 & 0.1063 & 0.1058 \\
\bottomrule
\end{tabular}
\label{tab:conv_wc}
\end{table}

\section{Variation of the two-band metric under a self-energy
correction}

Let $\mathbf{d}\to\mathbf{d}+\bm{\Sigma}(\mathbf{k})$. The normalized
vector changes by Eq.~\eqref{eq:deltad}, and the linear metric
correction is
\begin{equation}
\delta \metric_{ij}=\frac{1}{4}\left[
\partial_i\delta\hat{\mathbf{d}}\cdot\partial_j\hat{\mathbf{d}}
+\partial_i\hat{\mathbf{d}}\cdot\partial_j\delta\hat{\mathbf{d}}
\right],
\end{equation}
showing that only transverse self-energy components contribute at
leading order.

\bibliographystyle{apsrev4-2}
\bibliography{refs}

\end{document}